\documentclass{article}
\usepackage{graphicx}
\usepackage{epsfig}

\begin{document}
\begin{center}
 {\bf \Large Bargmann potentials and Marchenko transformations.}\\[3pt]
{\large  S.~A.~Zaitsev}\footnote{Supported in part by the Russian
Foundation of Basic Research under the grant No 02-02-17316 and
State Program Russian Universities}\\[2pt] {\sl Department of
Physics, Khabarovsk State Technical University,\\ Tikhookeanskaya
136, Khabarovsk 680035, Russia}\\[4pt]
\end{center}


\begin{abstract}
The Marchenko phase-equivalent transformation of the
Schr\"{o}dinger equation for two coupled channels is discussed.
The combination of the Marchenko transformations valid in the
Bargmann potential case is suggested.
\end{abstract}

\subsection*{Introduction}
\par Let the scattering data fit the necessary and sufficient conditions
(see, e.g. \cite{CS}) of the existence of the solutions of the
Marchenko equation
\begin{equation}
 \begin{array}{c}
\mathcal{K}_{\alpha \, \beta}(r, \, r')+\mathcal{Q}_{\alpha \, \beta}(r, \,
r')+ \int \limits _r^{\infty}\mathcal{K}_{\alpha \, \gamma}(r, \, t)\,
\mathcal{Q}_{\gamma \, \beta}(t, \, r')dt=0,\\[3mm]
\mathcal{Q}_{\alpha \, \beta}(r, \, r')= \frac{\displaystyle 1}{\displaystyle 2
\pi} \int \limits _{-\infty}^{\infty} dk [ \delta_{\alpha \, \beta}-S_{\alpha
\, \beta}] \,e^{\mbox{\scriptsize i}k(r+r')}+A_{\alpha}A_{\beta}e^{-\kappa \,
(r+r') }, \qquad \alpha,  \, \beta=1, \, 2.\\
 \end{array}
 \label{ME}
\end{equation}
Then the elements of resulting potential matrix
\begin{equation}
\left[\mathcal{V}_0(x)\right]_{\alpha \, \beta} = -2 \frac{\displaystyle
d}{\displaystyle d\,x}\mathcal{K}_{\alpha \, \beta}(x, \, x) \label{V0}
\end{equation}
for some $\varepsilon$ $(0 < \varepsilon <1)$ satisfy the condition \cite{AM}:
\begin{equation}
\int \limits _0 ^{\infty}\, x^{1+\theta}\, \left| \left[\mathcal{V}_0(x)
\right]_{\alpha\,\beta}\right|\,dx < \infty \qquad (-\varepsilon < \theta<
\varepsilon). \label{sumab}
\end{equation}
The purpose of the work is to construct the potential
(\ref{sumab}) which describes the interaction in a system of two
coupled channels of the angular momenta $\ell_1=0$ and $\ell_2=2$
starting from the Bargmann potential (\ref{V0}).

\par The transformations of the coupled Schr\"{o}dinger equation
\begin{equation}
{\Psi_0}''-\mathcal{V}_0(x)\,\Psi_0+ k^2 \Psi_0=0, \label{OrigE}
\end{equation}
into
\begin{equation}
{\Psi_1}''-\left(\mathcal{V}_1(x)+ \frac{\displaystyle 6}{\displaystyle
x^2}\,\mathcal{P}\right) \,\Psi_1+ k^2 \Psi_1=0, \quad
\mathcal{P}=\left(\begin{array}{cc}
0 & 0\\
0 & 1\\
\end{array}
\right)
\end{equation}
have been detailed in the book \cite{AM}, where the matrix
phase-equivalent transformations (MPET) associated with the
scattering data of the three different types have been suggested.
The conditions \cite{CS} corresponds to the case a) in chapter VII
of \cite{AM}, however the method of chapter VIII in Ref.
\cite{AM}) cannot be used in our case, since generally the
necessary condition for its implementation
\begin{equation}
\left[\mathcal{F}(0)\right]_{21} = 0, \label{ZCond}
\end{equation}
where $\mathcal{F}(k)$ is the Jost matrix
\begin{equation}
\left[\mathcal{F}(k)\right]_{\alpha\,\beta}=\mathcal{I}+\int
\limits_0^{\infty}\mathcal{K}_{\alpha \, \beta}(0, \, t)\,
e^{\mbox{\scriptsize i}kt}dt, \quad \mathcal{I}=\left(
\begin{array}{cc}
 1 & 0\\
 0 & 1\\
\end{array} \right), \label{JM}
\end{equation}
does not hold. As a consequence $\mathcal{V}_1(x)$ has the singularity $x^{-2}$
at the origin or at infinity.
\par Below are given the auxiliary (direct and inverse) Marchenko
transformations which in combination with MPET furnish the potential which is
regular at the origin and decreases at infinity faster than $x^{-2}$.

\subsection*{Bargmann potential properties}
\par We proceed from the assumption that the elements of the potential matrix
(\ref{V0}) differ from zero at the origin:
\begin{equation}
\left[\mathcal{V}_0(0)\right]_{\alpha, \, \beta} \neq 0. \label{V_0}
\end{equation}
In that case it can been shown that the regular solution $\mathcal{G}_0(x, \,
k)$ of Eq. (\ref{OrigE}) has the behavior near the origin:
\begin{equation}
\mathcal{G}_0(x, \, k)= \left(
 \begin{array}{cc}
  x + \lambda \, x^3+o(x^3) & \nu \, x^3+o(x^3)\\[3mm]
  \nu \, x^3+ o(x^3) & x +\mu \, x^3+o(x^3)\\
 \end{array}
 \right), \quad (x \to 0),
 \label{RegSolOrig}
\end{equation}
where
\begin{equation}
\lambda = \frac{[\mathcal{V}_0(0)]_{11}-k^2}{6}, \; \mu =
\frac{[\mathcal{V}_0(0)]_{22}-k^2}{6}, \; \nu
=\frac{[\mathcal{V}_0(0)]_{12}}{6}.
 \label{V0Orig}
\end{equation}
The asymptotic behavior of $\mathcal{G}_0(x, \, \mbox{i}\sigma)$, $\sigma \geq
0$ is given by the following expression \cite{AM}:
\begin{equation}
\mathcal{G}_0 \rightarrow
\frac{e^{\sigma\,x}}{2\,\sigma}\,\widetilde{\mathcal{F}}(\mbox{i}\sigma), \;
(\sigma >0); \qquad \mathcal{G}_0 \rightarrow x \,\widetilde{\mathcal{F}}(0) +
o(x^{1-\theta}), \; (\sigma=0),
\end{equation}
where tilde denotes transposition.

\par In the following, the functions
\begin{equation}
\mathcal{Y}_0(x) = \mathcal{G}_0(x, \, \mbox{i} \sigma)\,\mathcal{P}, \; \sigma
> 0, \quad \mathcal{Z}_0(x) = \mathcal{G}_0(x, \, 0)\,\mathcal{P}
\end{equation}
are used. We also will keep track of the bound state wave function $\Phi_0(x)
\equiv \Psi_0(x, \, \mbox{i}\kappa)$, $\kappa
> 0$:
\begin{equation}
\Phi_0(x)= \left(
 \begin{array}{c}
  u_1 \, x + u_2 \, x^3 + o(x^3)\\[3mm]
  v_1 \, x + v_2 \, x^3 + o(x^3)\\
 \end{array}
 \right), \quad (x \to 0),
 \label{BoundWF0}
\end{equation}
\begin{equation}
\Phi_0(x) \rightarrow e^{-\kappa \, x} \left(
 \begin{array}{c}
  A_1\\
  A_2\\
 \end{array}
 \right), \quad (x \to \infty)
 \label{BoundWFinf}
\end{equation}
transformations.

\subsection*{First transformation} The intent of this transformation is to
obtain the Jost function that satisfies the condition (\ref{ZCond}). Here, the
matrix-function
\begin{equation}
\mathcal{Y}_1=\mathcal{Y}_0\,\mathcal{J}_1^{-1},\quad
\mathcal{J}_1=\mathcal{I}+ \int \limits _0 ^x
\widetilde{\mathcal{Y}_0}(t)\,\mathcal{Y}_0(t)\,dt
\end{equation}
is constructed in such a way that the matrix
\begin{equation}
\Psi_1(x, \, k)=\Psi_0(x, \, k)+\mathcal{Y}_1(x)\, \mathcal{W}\left\{
\widetilde{\mathcal{Y}_0}(x), \; \Psi_0(x, \, k)\right\}\,(\sigma^2+
k^2)^{-1}, \qquad (k \neq \mbox{i}\sigma) \label{Psi1}
\end{equation}
obeys the Schr\"{o}dinger equation
\begin{equation}
{\Psi_1}''-\mathcal{V}_1(x)\,\Psi_1+ k^2 \Psi_1=0 \label{FTE}
\end{equation}
with the transformed potential $\mathcal{V}_1(x)$:
\begin{equation}
\mathcal{V}_1(x)=\mathcal{V}_0(x)-2\left[\mathcal{Y}_1(x)\,
\widetilde{\mathcal{Y}_0}(x)\right]'. \label{V1}
\end{equation}
Here, $\mathcal{W}\left\{ \widetilde{\mathcal{X}}(x), \; \mathcal{Y}(x)
\right\}$ denotes the Wronskian: $\mathcal{W}\left\{ \widetilde{\mathcal{X}},
\; \mathcal{Y} \right\}$=$\widetilde{\mathcal{X}}\,{\mathcal{Y}}\,'$-
${\widetilde{\mathcal{X}}}\,'\,\mathcal{Y}$.

\par The asymptotic behavior of the functions $\mathcal{Y}_0$ and
$\mathcal{Y}_1(x)$ \cite{AM}:
\begin{equation}
\mathcal{Y}_0 \rightarrow e^{\sigma\,x} \,\mathcal{C}\,\mathcal{P}, \quad
\mathcal{C} = \frac{1}{2\, \sigma}\, \widetilde{\mathcal{F}}(\mbox{i}\sigma),
\end{equation}
\begin{equation}
\mathcal{Y}_1 \rightarrow
2\sigma\,s^{-1}e^{-\sigma\,x}\,\mathcal{C}\,\mathcal{P}, \quad s \,
\mathcal{P} = \mathcal{P}\,\widetilde{\mathcal{C}}\,\mathcal{C} \,\mathcal{P},
\end{equation}
provides the modification of the initial $S$ matrix and Jost matrix \cite{AM}:
\begin{equation}
\widehat{\mathcal{S}}(k) = \mathcal{T}(k)\,\mathcal{S}(k)\, \mathcal{T}(k),
\quad \widehat{\mathcal{F}}(k)=\mathcal{F}(k)\,\mathcal{T}(k), \label{TSF}
\end{equation}
where
\begin{equation}
\mathcal{T}(k)= \mathcal{I}-2\sigma\,(\sigma+\mbox{i}k)^{-1}s^{-1}\mathcal{C}\,
\mathcal{P}\,\widetilde{\mathcal{C}}.
\end{equation}
The parameter $\sigma$ value is defined by the expression:
\begin{equation}
\left[\mathcal{F}(0) \,\mathcal{T}(0)\right]_{21}=0,
\end{equation}
that is equivalent to
\begin{equation}
\zeta^2-2\zeta_0\zeta-1=0, \label{zeta}
\end{equation}
where
$\zeta(\sigma)=\mathcal{F}_{22}(\mbox{i}\sigma)/\mathcal{F}_{21}(\mbox{i}\sigma)$,
$\zeta_0=\zeta(0)$.

\par The solution $\mathcal{Z}_1(x)$ of Eq. (\ref{FTE}) asymptotic behavior is
given by the expression:
\begin{equation}
\mathcal{Z}_1(x) \rightarrow \left(\left[\widehat{\mathcal{F}}(0)\right]_{22}\,
x \, \mathcal{I}+ o(x^{1-\theta}) \right) \mathcal{P}, \quad (x \rightarrow
\infty). \label{AsZ1}
\end{equation}

\par One can see from the regular solution $\Phi_0(x, \, k)$ of Eq.
(\ref{OrigE}) transformation
\begin{equation}
\Phi_1(x, \, k) = \Phi_0(x, \, k) - \mathcal{Y}_1(x)\int \limits _0
^{x}\widetilde{\mathcal{Y}}_0(t)\,\Phi_0(t, \, k)\,dt
\end{equation}
that $\Phi_0(x, \, k)$ and $\Phi_1(x, \, k)$ have the same behavior at the
origin and $\mathcal{V}_1(0)=\mathcal{V}_0(0)$.
\subsection*{Second transformation}
\par Now that the Jost function $\widehat{\mathcal{F}}(k)$ meets the condition
$\left[\widehat{\mathcal{F}}(0)\right]_{21} = 0$, MPET \cite{AM} may be
implemented to Eq. (\ref{FTE}). Here, the solution $\mathcal{Z}_1(x)$ of Eq.
(\ref{FTE}) and matrix-function
\begin{equation}
\mathcal{Z}_2=\mathcal{Z}_1\,\mathcal{J}_2^{-1}, \quad \mathcal{J}_2
=\mathcal{I}-\mathcal{P}-\int \limits _0 ^x
\widetilde{\mathcal{Z}_1}(t)\,\mathcal{Z}_1(t)\,dt \label{Z2}
\end{equation}
are used for the transformation construction \cite{AM}:
\begin{equation}
\Psi_2(x, \, k)=\Psi_1(x, \, k)-\mathcal{Z}_2(x)\, \mathcal{W}\left\{
\widetilde{\mathcal{Z}_1}(x), \; \Psi_1(x, \, k)\right\}\,k^{-2}.
\end{equation}
$\Psi_2(x, \, k)$ obeys the equation
\begin{equation}
{\Psi_2}''-\left(\mathcal{V}_2(x)+ \frac{\displaystyle 6}{\displaystyle
x^2}\,\mathcal{P}\right) \,\Psi_2+ k^2 \Psi_2=0, \label{STE}
\end{equation}
with the potential:
\begin{equation}
\mathcal{V}_2(x)=\mathcal{V}_1(x)+2\left[\mathcal{Z}_2(x)\,
\widetilde{\mathcal{Z}_1}(x)\right]'-\frac{\displaystyle 6}{\displaystyle
x^2}\,\mathcal{P}. \label{V2}
\end{equation}

\par The behavior of $\Phi_2(x)$ and $\mathcal{Y}_2(x)$ at the origin:
\begin{equation}
\begin{array}{c}
 \Phi_2(x)= \left(
 \begin{array}{c}
  u_1 \, x + {u_2}' \, x^3 + o(x^3)\\[3mm]
  {v_2}' \, x^3 + o(x^3)\\
 \end{array}
 \right), \quad
\mathcal{Y}_2(x)= \left(
 \begin{array}{cc}
  0 & o(x^3)\\[3mm]
  0 & {\mu}' \, x^3 + o(x^3)\\
 \end{array}
 \right),\\
 \end{array}
 \label{Orig2}
\end{equation}
can be deduced from the regular solution transformation formula \cite{AM}:
\begin{equation}
\Phi_2(x, \, k) = \Phi_1(x, \, k) + \mathcal{Z}_2(x)\int \limits _0
^{x}\widetilde{\mathcal{Z}}_1(t)\,\Phi_1(t, \, k)\,dt.
\end{equation}
Reference to Eq. (\ref{Z2}) shows that
\begin{equation}
\mathcal{Z}_2(x)= \left(
 \begin{array}{cc}
  0 & -3\, \nu +o(1)\\[3mm]
  0 & -\frac{\displaystyle 3}{\displaystyle x^2} +O(1)\\
 \end{array}
 \right), \quad (x \to 0),\\
\end{equation}
so that the second term in the right-hand side of (\ref{V2}) reads
\begin{equation}
2\left[\mathcal{Z}_2(x)\, \widetilde{\mathcal{Z}_1}(x)\right]'= \left(
\begin{array}{cc}
 -18\, \nu^2\, x^2 + o(x^2) & -6\, \nu + o(x)\\[3mm]
  -6\, \nu + o(x) & \frac{\displaystyle 6}{\displaystyle x^2}+ O(1)\\
\end{array}
 \right), \quad (x \to 0).
\label{Add2Orig}
\end{equation}
From (\ref{Add2Orig}) it follows that $\mathcal{V}_2(x)$ is regular at the
origin and $\left[\mathcal{V}_2(0) \right]_{12}=0$.

\par $\mathcal{Z}_2(x)$ behaves at infinity like \cite{AM}
\begin{equation}
\mathcal{Z}_2(x) \rightarrow \left(-3\,
\left[\widehat{\mathcal{F}}(0)\right]_{22}^{-1}\, x^{-2} \mathcal{I}+
o(x^{-2-\theta}) \right) \, \mathcal{P}, \quad (x \to \infty).  \label{AsZ2}
\end{equation}
It follows from (\ref{AsZ1}) and (\ref{AsZ2}) that \cite{AM}
\begin{equation}
2 \left[ \mathcal{Z}_2(x) \, \widetilde{\mathcal{Z}}_1(x)\right]' \rightarrow
\left(\frac{\displaystyle 6}{\displaystyle x^2}\, \mathcal{I}+ o(x^{-2-\theta})
\right)\, \mathcal{P}, \quad (x \to \infty). \label{Add2As}
\end{equation}
One can see from (\ref{STE}), (\ref{V2}), (\ref{Add2Orig}) and (\ref{Add2As})
that MPET increases the second channel angular momentum to 2.

\subsection*{Third transformation}
\par Consider the following transformation:
\begin{equation}
 \begin{array}{c}
\Psi_3(x, \, k)=\Psi_2(x, \, k)+\mathcal{Y}_3(x)\, \mathcal{W}\left\{
\widetilde{\mathcal{Y}_2}(x), \; \Psi_0(x, \,
k)\right\}\,(\sigma^2+k^2)^{-1},\\[3mm]
\mathcal{V}_3(x)=\mathcal{V}_2(x)-2\left[\mathcal{Y}_3(x)\,
\widetilde{\mathcal{Y}_2}(x)\right]', \\
 \end{array}
\label{V3}
\end{equation}
where
\begin{equation}
\mathcal{Y}_3=\mathcal{Y}_2\,\mathcal{J}_3^{-1}, \quad
\mathcal{J}_3=\mathcal{I}-\mathcal{P}- \int \limits _x ^{\infty}
\widetilde{\mathcal{Y}_2}(t)\,\mathcal{Y}_2(t)\,dt.
\end{equation}
From theorem $6.2.1$ in the book \cite{AM} it follows that this transformation
is asymptotically inverse of the first one, since $\mathcal{Y}_3(x)=$ $
-\mathcal{Y}_0(x)$ and $-\mathcal{J}_3(x)=\mathcal{J}^{-1}_1(x)$ at infinity.

\par Notice that the condition (\ref{sumab}) is not affected by each of the three
transformations.

\subsection*{Other angular momenta} Note that the regular solution
$\mathcal{G}_3(x)$ of
\begin{equation} {\mathcal{G}_3}''-\left(\mathcal{V}_3(x)+
\frac{\displaystyle 6}{\displaystyle x^2}\,\mathcal{P}\right) \,\mathcal{G}_3=0
\label{TTE}
\end{equation}
then can be used to equally increase the angular momenta in both channels.
Indeed, MPET corresponding to the transformations \cite{M}, can be written in
the forms:
\begin{equation}
 \begin{array}{c}
\Psi_4(k, \, x)=-k^{-2} \widetilde{\mathcal{G}}_3^{-1}(x)\,W
\left\{\widetilde{\mathcal{G}}_3(x), \; \Psi_3(k, \, x) \right\},\\[3mm]
\mathcal{V}_4(x) = \mathcal{V}_3(x) -2 \left[\widetilde{\mathcal{G}_3}^{-1}(x)
\, \widetilde{\mathcal{G}_3}'(x)\right]';\\
 \end{array}
 \label{V4a}
\end{equation}
\begin{equation}
 \begin{array}{c}
\Psi_4(k, \, x)=\Psi_3(k,\, x)+k^{-2} \mathcal{X}(r)\,W
\left\{\widetilde{\mathcal{G}}_3(x), \; \Psi_3(k, \, r) \right\},\\[3mm]
\mathcal{V}_4(x) = \mathcal{V}_3(x) -2 \left[\mathcal{X}(x) \,
\widetilde{\mathcal{G}}_3(x)\right]',\\
 \end{array}
 \label{V4b}
\end{equation}
where
\begin{equation}
\mathcal{X}(x)=\mathcal{G}_3(x) \left[\int \limits _0 ^x
\widetilde{\mathcal{G}}_3(t) \, \mathcal{G}_3(t) \, dt \right]^{-1}. \label{X}
\end{equation}
MPET (\ref{V4a}) and (\ref{V4b}) increase each of $\ell_1$, $\ell_2$ by 1 and
2, correspondingly, provided that ${\mathcal{G}_3}^{-1}(x)$ and
$\mathcal{X}(x)$ exist.

\subsection*{Example}
\par As an example the potential is calculated corresponding the simplest
$\mathcal{S}$ matrix coupling S and D states \cite{Newt1}:
\begin{equation}
\mathcal{S}(k)=\frac{\displaystyle 1}{\displaystyle k^4+4\chi^4}
 \left(
  \begin{array}{cc}
   2\chi^2 & k^2\\[2mm]
   -k^2    & 2\chi^2\\
  \end{array}
 \right)
  \left(
  \begin{array}{cc}
   \left( \frac{\displaystyle k+\mbox{i}\phi}{\displaystyle k-\mbox{i}\phi}\right)
   \left( \frac{\displaystyle k+\mbox{i}\kappa}{\displaystyle k-\mbox{i}\kappa}
    \right) & 0\\[2mm]
   0       & 1\\
  \end{array}
 \right)
 \left(
  \begin{array}{cc}
   2\chi^2 & -k^2\\[2mm]
   k^2    & 2\chi^2\\
  \end{array}
 \right).
 \label{NS}
\end{equation}
\par Notice that the transformation $\mathcal{V}_0$ $\rightarrow$ $\mathcal{V}_3$
have been carried out using MPET, i. e. the solution $\Psi_3(x, \, k)$, $(k >
0)$ has asymptotic behavior identical to that of $\Psi_0(x, \, k)$:
\begin{equation}
\Psi_0(x, \, k)\rightarrow \frac{\mbox{\footnotesize i}}{\mbox{\footnotesize
2}} \left\{ e^{-\mbox{\scriptsize i} k\,x} \mathcal{I}- e^{\mbox{\scriptsize i}
k\,x} \mathcal{S}\right\}, \quad (x \rightarrow \infty).
\end{equation}
Thus in order that the asymptotic properties of $\Psi_3(x, \, k)$ correspond
to the potential $\mathcal{V}_3$ \cite{Newton}:
\begin{equation}
\Psi_3(x, \, k)\rightarrow \frac{\mbox{\small i}}{\mbox{\footnotesize 2}}\,
e^{\frac{\mbox{\tiny i}}{2} \pi\,\mathcal{L}}\left\{ e^{-\mbox{\scriptsize i}
k\,x} \mathcal{I}- e^{-\mbox{\scriptsize i}
\pi\,\mathcal{L}}\,e^{\mbox{\scriptsize i} k\,x} \mathcal{S}\right\}, \quad (x
\rightarrow \infty),
\end{equation}
where
\begin{equation}
\mathcal{L}= \left( \begin{array}{cc}
                     0 & 0\\
                     0 & 2\\
                    \end{array}
                    \right),
\end{equation}
the matrix
\begin{equation}
\overline{\mathcal{S}}=e^{-\frac{\mbox{\tiny i}}{2} \pi\,\mathcal{L}}\,
\mathcal{S}\,e^{-\frac{\mbox{\tiny i}}{2} \pi\,\mathcal{L}}
\end{equation}
must be used as $\mathcal{S}$ matrix in the Marchenko equations (\ref{ME}). In
this instance (\ref{NS}) is used as $\overline{\mathcal{S}}$, since, for
example, all values $-\mbox{i}\mathop{Res} \limits _{k = \mbox{\scriptsize
i}\kappa}S_{\alpha, \, \beta}(k)$ are negative and hence the corresponding
asymptotic normalization constants $A_1$, $A_2$ are positive.
\par
\begin{figure}
\centerline{\psfig{figure=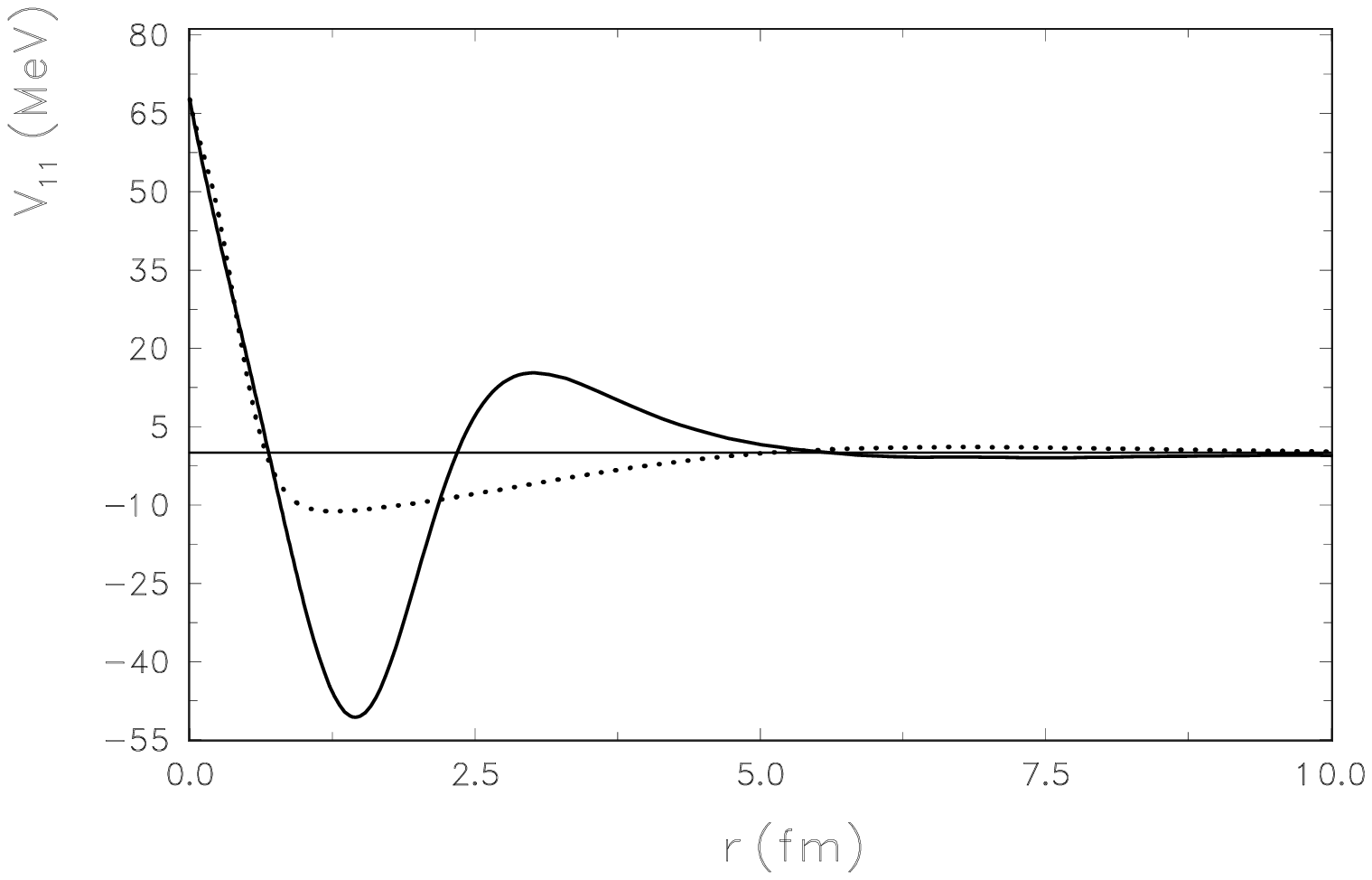,width=1.\textwidth}}
\caption{}\label{V11}
\centerline{\psfig{figure=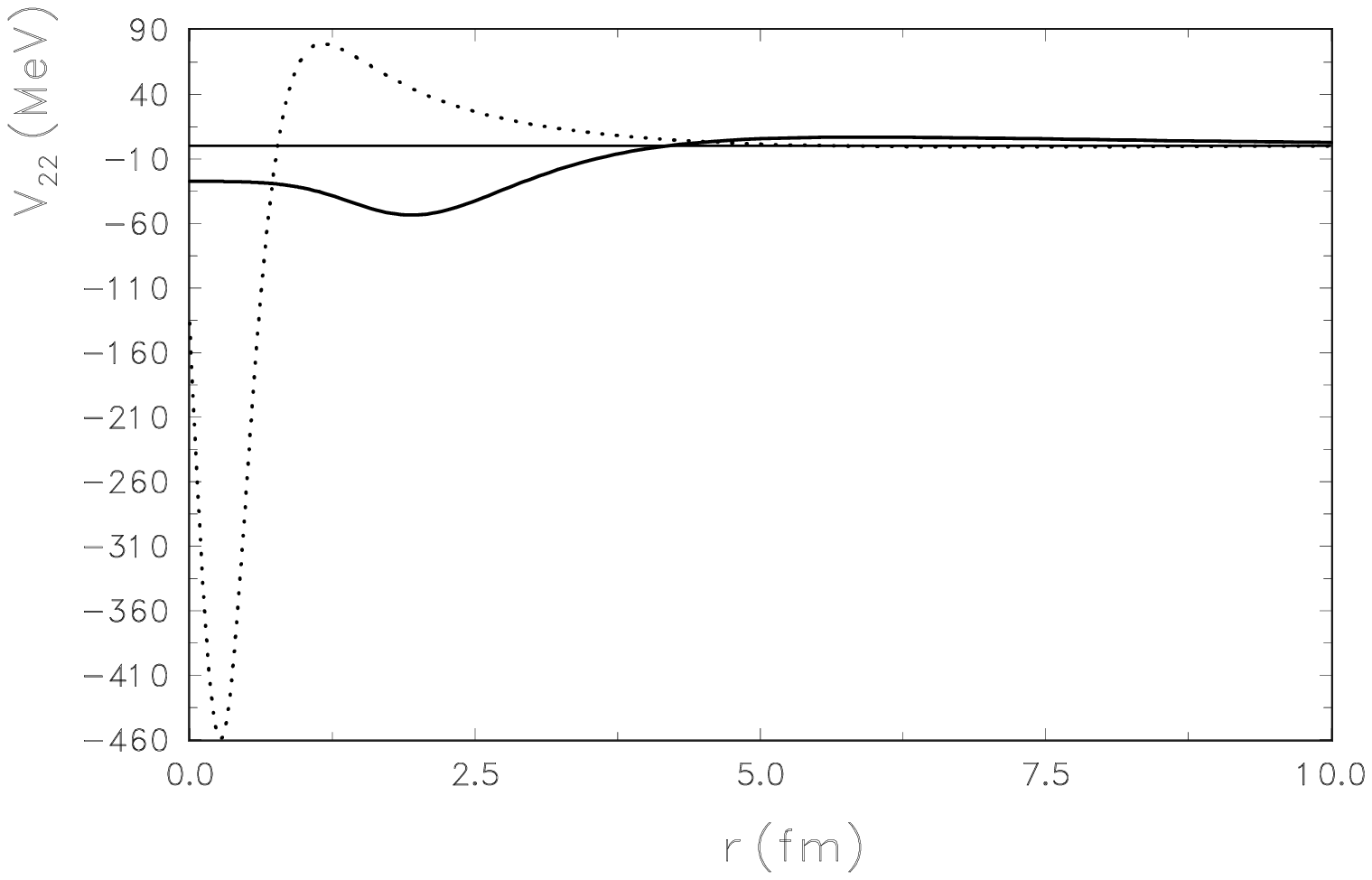,width=1.\textwidth}}
\caption{}\label{V22}
\end{figure}

\begin{figure}
\centerline{\psfig{figure=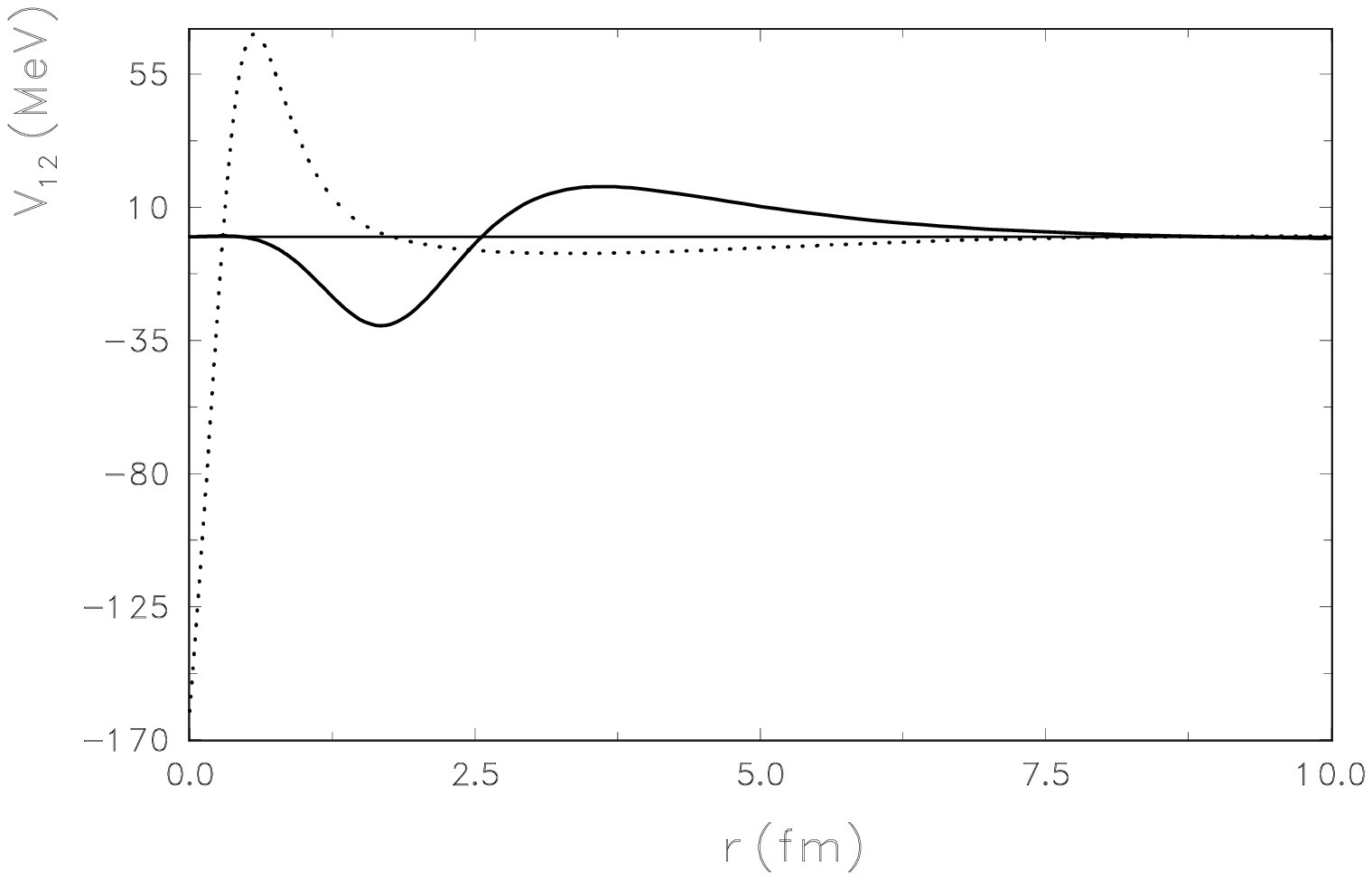,width=1.\textwidth}} \caption{}
\label{V12}
\centerline{\psfig{figure=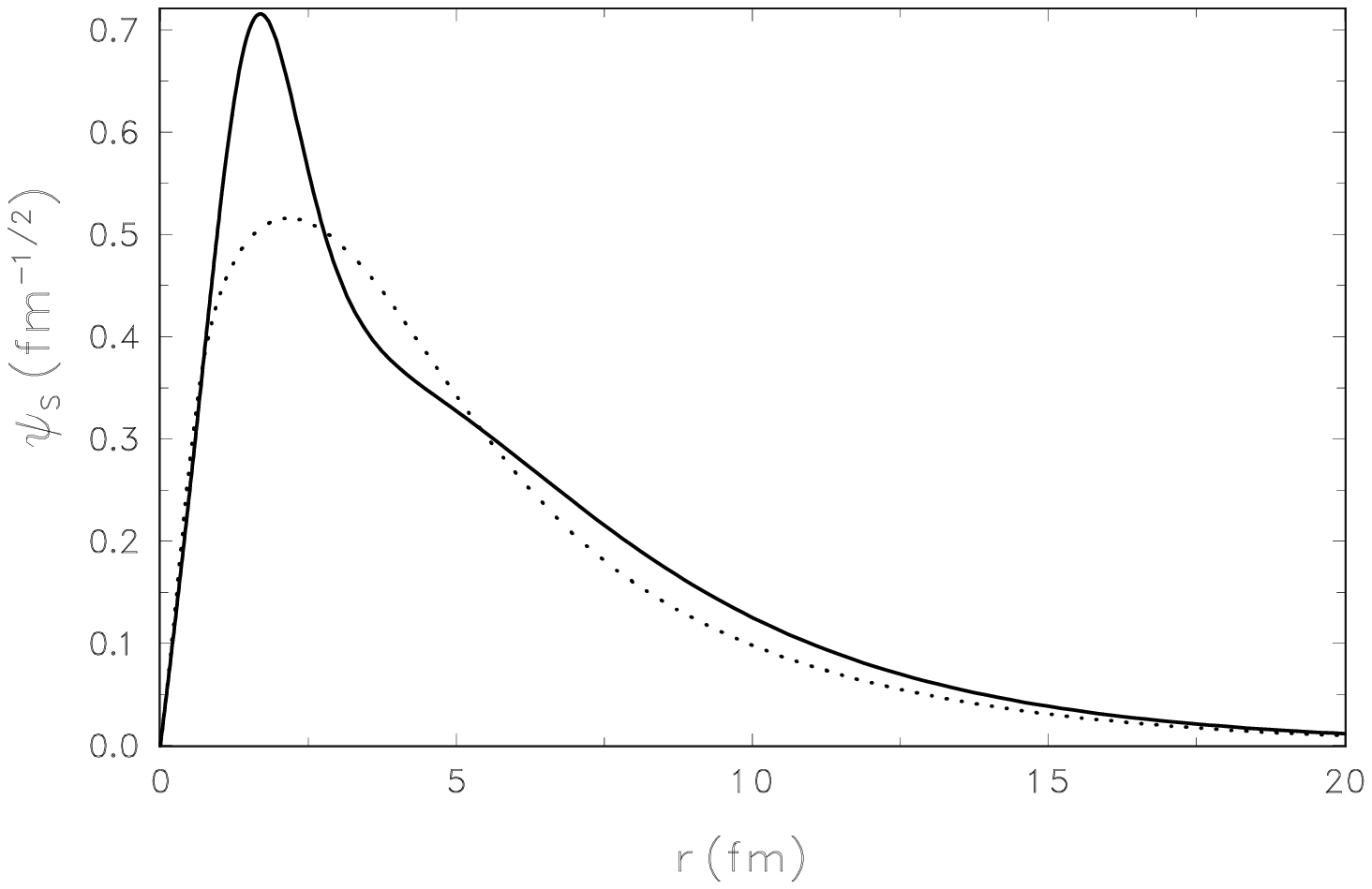,width=1.\textwidth}}
\caption{}\label{Swf}
\end{figure}

\begin{figure}
\centerline{\psfig{figure=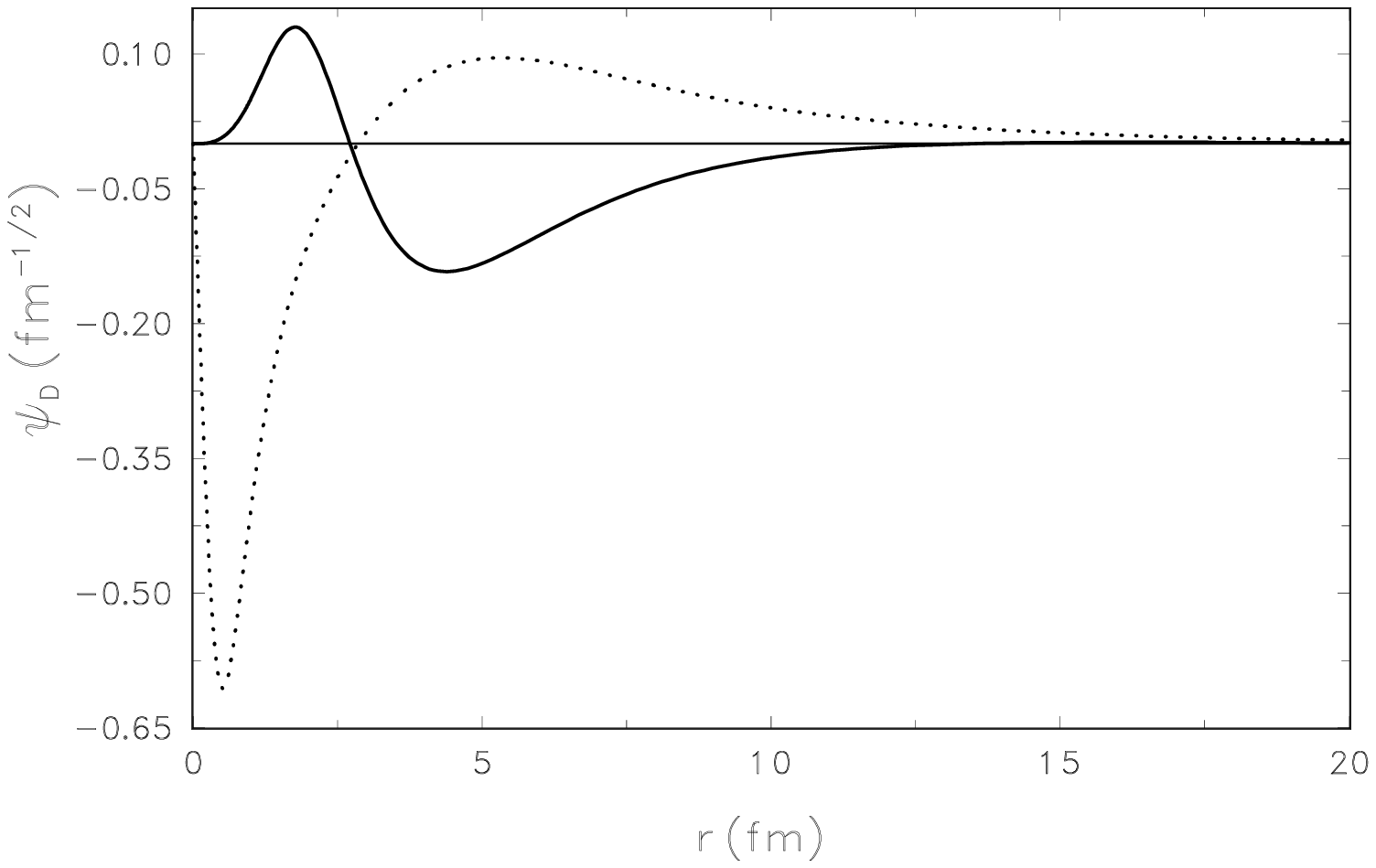,width=1.\textwidth}}
\caption{} \label{Dwf}
\centerline{\psfig{figure=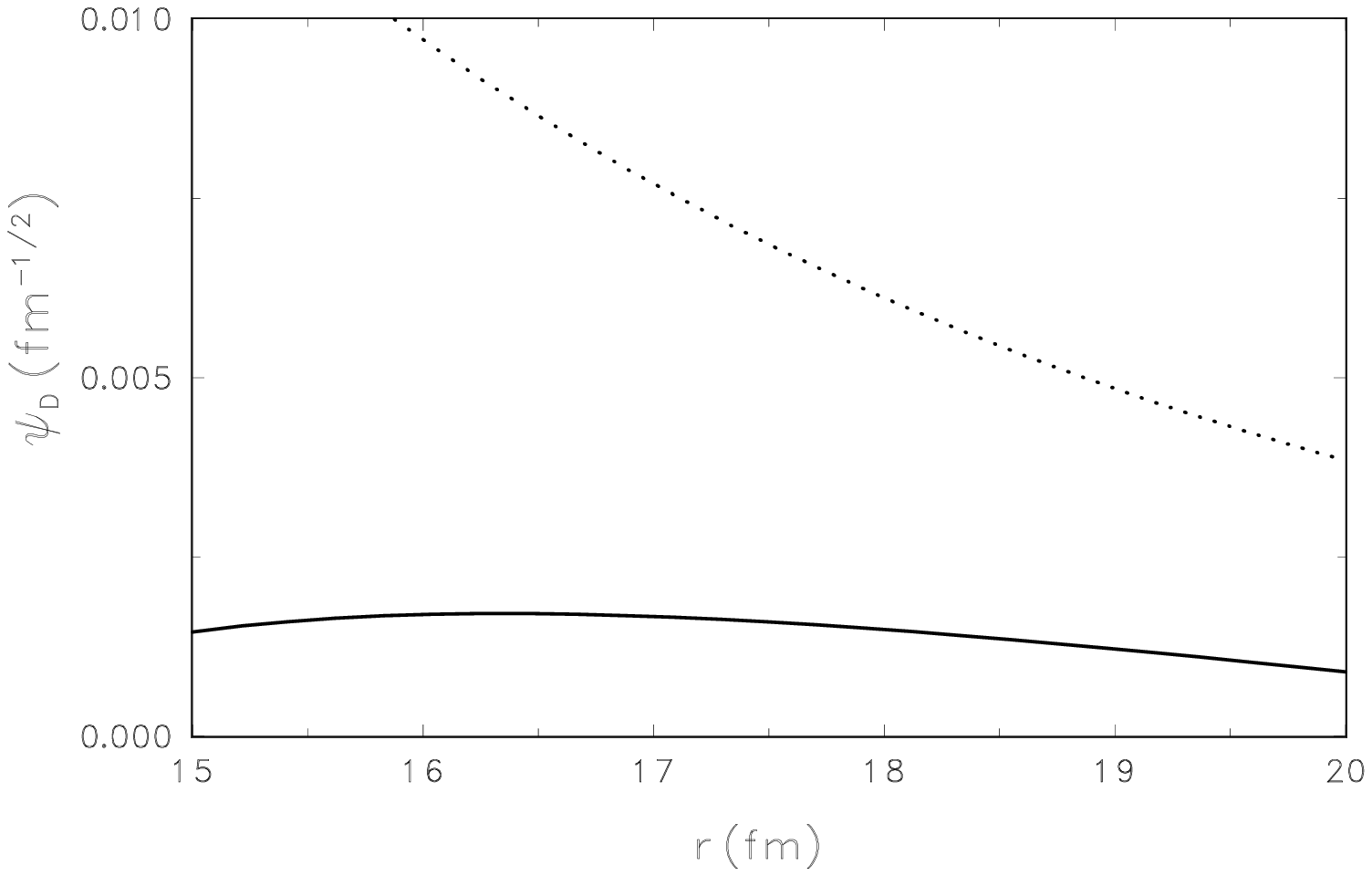,width=1.\textwidth}}
\caption{} \label{Dwfa}
\end{figure}

 For the parameters $\chi=0.26$ $fm^{-1}$, $\phi=0.944$ $fm^{-1}$ and $\kappa=0.232$
$fm^{-1}$ \cite{Newt1} Eq. (\ref{zeta}) gives $\sigma=0.2053483144$ $fm^{-1}$.
The initial Bargmann potentials $\left[\mathcal{V}_0(x)\right]_{\alpha, \,
\beta}$ and resulting ones $\left[\mathcal{V}_3(x)\right]_{\alpha, \, \beta}$
are presented in Figures \ref{V11}-\ref{V12} by dotted and solid lines
correspondingly. Figures \ref{Swf}, \ref{Dwf} exhibit the S and D components
of the bound state (with the energy $-\kappa^2$) wave functions: $\Phi_0(x)$
(dotted line), $\Phi_3(x)$ (solid line). Note that $\Phi_3(x)$ in the interior
domain may be modified by, for example, the multichannel phase-equivalent
transformation \cite{ShMaz}. The asymptotic forms of $\Phi_0(x)$ and
$\Phi_3(x)$ D components are represented by dotted and solid lines in Fig.
\ref{Dwfa}.

\newpage
\clearpage

\newpage
\clearpage

\newpage
\clearpage


\begin{thebibliography}{99}

\bibitem{CS}
 K.~Chadan and P.~C.~Sabatier. {\sl Inverse problem in quantum scattering theory}
 (2nd ed., Springer-verlag, New York, 1989.)

\bibitem{AM}
 Z.~S.~Agranovich and V.~A.~Marchenko, {\sl The Inverse Problem of the Scattering
 Theory} (Gordon and Breach, New York, 1963).

\bibitem{M}
V.~A. Marchenko. {\sl Sturm-Liouville operators and applications} (Berkhauser,
Basel, 1986).

\bibitem{Newton}
R.~G.~Newton. {\sl Scattering Theory of Waves and Particles} (McGraw-Hill, New
York, 1966.)

\bibitem{Newt1}
R.~Newton, T.~Fulton. Phys. Rev. {\bf 107}, 1103, 1957.

\bibitem{ShMaz}
A.~M.~Shirokov, V.~N.~Sidorenko.
Phys.Atom.Nucl. 63 (2000) 1993; Yad.Fiz. 63 (2000) 2085.



\end{thebibliography}
\end{document}